# بررسی مقاومت الگوریتم مجموعه عضوی انعکاس آفین

رجب شعبانی

دانشگاه پیام نور

**Abstract**

In this letter, we study the local and the global robustness of the set-membership affine projection (SM-AP) algorithm. We demonstrate that the SM-AP algorithm has $l_2$-stability. In fact, the SM-AP algorithm never diverges; no matter how the parameters of the SM-AP algorithm has been adopted. Finally, the simulation results substantiate the validity of the proposed analysis.

چکیده

در این مقاله، دو ویژگی برای مقاومت الگوریتم مجموعه عضوی انعکاس آفین (SM-AP) ارائه داده می شود. در ویژگی اول یک کران موضعی برای میزان خطا/اختلاف ضرایب در هر تکرار ارائه می گردد، اما ویژگی دوم این خاصیت را برای هر تکرار بسط می دهد. در نتیجه در این مقاله برای اولین بار نشان می دهیم که الگوریتم SM-AP فارغ از چگونگی گزینش پارامترها از جهت پایداری با نرم $l_2$ مقاوم است.

کلید واژه ها: فیلترهای تطبیقی، مجموعه عضوی، انعکاس آفین، مقاومت، پایداری با نرم $l_2$، کران خطا

## ۱   مقدمه

فیلترهای تطبیقی کلاسیک روش های تخمین بازگشتی هستند که بر اساس تقریب نقطه ای کار می کنند[۱]. این تکنیک به دنبال یک نقطه خاص به عنوان جواب برای مساله بهینه سازی است. دو الگوریتم مهم این روش الگوریتم مترین میانگین مربعات نرمال شده ( NLMS ) و الگوریتم انعکاس آفین ( AP )هستند. این دو الگوریتم در کمترین میزان خطا و بالاترین سرعت همگرایی با هم در تقابل هستند[۲، ۳، ۴].

۱

اما الگوریتم های اندکی در فیلترهای تطبیقی هستند که از روش مجموعه عضوی استفاده می کنند [۵]. از جمله این الگوریتم ها، فیلترهای تطبیقی مجموعه عضوی (SMF)) هستند. در روش مجموعه عضوی یک مجموعه جواب $\Theta$ تعریف می شود و هر عضو این مجموعه مورد قبول واقع می شود. چون مسایل در دنیای واقعی دارای نا اطمینانی های فراوانی هستند، معقولانه تر است به جای یک جواب به کمک روش مجموعه عضوی به دنبال مجموعه ای از جواب ها باشیم.

فیلترهای تطبیقی مجموعه عضوی روش مجموعه عضوی را با روش انتخاب داده ترکیب کرده و الگوریتم های مجموعه عضوی را ارائه می دهند [۶، ۷، ۸]. روش انتخاب داده مسئول هزینه محاسباتی الگوریتم های مجموعه عضوی است. در واقع تنها اگر میزان خطای الگوریتم بزرگتر از یک کران از پیش تعیین شده باشد، پارامترهای الگوریتم بروزرسانی می گردد [۹، ۱۰، ۱۱]. یکی از الگوریتم های مجموعه عضوی مهم الگوریتم مجموعه عضوی انعکاس آفین (SM-AP) می باشد [۱۲]. الگوریتم SM-AP با حفظ مزیت های الگوریتم AP، به کمک خاصیت داده-انتخابی می تواند قبل از استفاده از داده های جدید آنها را ارزیابی کند و در صورت مفید بودن آنها را در فرایند یادگیری مورد استفاده قرار دهد. بدین وسیله الگوریتم SM-AP در مقایسه با AP دارای دقت بیشتر و مقاومت بیشتر در برابر نویز است و همچنین هزینه محاسباتی کمتری را برای بروزرسانی ضرایب صرف می کند زیرا تنها در صورتی داده های جدید را در فرایند یادگیری به کار می گیرد که در آنها اطلاعات جدیدی موجود باشد [۱۳، ۱۴، ۱۵، ۱۶]. الگوریتم مجموعه عضوی فراوانی در صنعت مورد استفاده قرار گرفته اند، از جمله آنها می توان به مقالات [۱۷، ۱۸، ۱۹، ۲۰] مراجعه کرد.

با وجود تمام مزیت هایی که برای الگوریتم SM-AP وجود دارد، اما این الگوریتم هنوزه در صنعت به اندازه کافی مورد استفاده قرار نمی گیرد، دلیل آن می تواند کمبود بررسی ویژگی های تئوری این الگوریتم باشد [۱۵]. پایداری الگوریتم SM-NLMS از نظر نرم l2 در [۱۴، ۲۱] مورد بررسی قرار گرفته است. در این مقاله، پایداری الگوریتم SM-AP را از نظر نرم l1 مورد بررسی قرار می دهیم. در بخش ۲ کمی در مورد پایداری الگوریتم بحث می کنیم. در بخش ۳ الگوریتم SM-AP را مرور می کنیم. سپس در بخش ۴ دو خاصیت کلی را برای میزان پایداری این الگوریتم مورد بحث قرار می دهیم. در بخش ۵ شبیه سازی و آزمایش ها را ارائه می کنیم و در انتها نتیجه گیری در بخش ۶ جمع بندی می شود.

نمادها: برای تکرار k، ضرایب فیلتر تطبیقی، فیلتر بهینه، تفاضل بین فیلتر تطبیقی و فیلتر بهینه و بردار ورودی به ترتیب با $w(k)$، $w_o$، $\widetilde{w}(k)$ و $x(k)$ در $R^{N+1}$ نشان داده می شوند. سیگنال های مرجع، خروجی، خطا و نویز به ترتیب با $d(k), y(k)$، $e(k)$ و $n(k)$ در R نمایش داده می شوند که $y(k) = x^T(k)w(k)$ و $e(k) = d(k) - y(k)$. ماتریس ورودی، بردار مرجع و بردار خطا با $X(k)$، $d(k)$ و $e(k)$ نمایش داده می شوند.

۲

## ۲  شرط پایداری

برای تکرار k، فرض کنید که سیگنال مرجع (d(k از طریق رابطه زیر با سیستم ناشناخته $w_o$ مرتبط شده است

$$d(k) \triangleq \underbrace{w_o^T x(k)}_{\triangleq y_o(k)} + n(k), \qquad (1)$$

که (n(k سیگنال نویز است و ضریب عدم اطمینان و خطا را مدل سازی می کند. همچنین فرض می کنیم که دنباله نویز {(n(k} دارای انرژی متناهی است [۳]، یعنی

$$\sum_{k=0}^{j} |n(k)|^2 < \infty, \qquad \forall\, j. \qquad (2)$$

فرض کنید که دنباله سیگنال مرجع {(d(k} را داریم و می خواهیم $y_o(k) = w_o^T x(k)$ را تخمین بزنیم. در این راستا فرض کنید $\hat{y}_{k|k}$ تخمینی از $y_o(k)$ است و تنها به ازای $j = 0, \cdots, k$ وابسته است. برای مقدار ثابت $\eta$، می خواهیم تخمین های $\hat{y}_{k|k} \in \{\hat{y}_{0|0}, \hat{y}_{1|1}, \cdots, \hat{y}_{N|N}\}$ را به گونه ای حساب کنیم که برای هر (n(k که در (۲) صدق می کند، رابطه زیر برقرار باشد:

$$\frac{\sum_{k=0}^{j} ||\hat{y}_{k|k} - y_o(k)||^2}{\widetilde{w}^T(0)\widetilde{w}(0) + \sum_{k=0}^{j} |n(k)|^2} < \eta^2, \quad \forall\, j = 0, \cdots, N \qquad (3)$$

که در آن $\widetilde{w}(0) \triangleq w_o - w(0)$ و (w(0 مقدار حدس اولیه $w_o$ است. توجه داشته باشید که صورت تخمینی از میزان انرژی خطا تا تکرار k و مخرج شامل انرژی عدم اطمینان تا تکرار k و انرژی خطا (0)$\widetilde{w}$ است.

در نتیجه معیار (۳) ما را ملزم می کند که تخمین های {$\hat{y}_{k|k}$} را به گونه ای حساب کنیم که نسبت تخمین انرژی خطا (صورت) به انرژی عدم اطمینان (مخرج) از $\eta^2$ فراتر نرود. زمانی که این معیار صدق می کند، می گوییم که تخمین به دست آمده مقاوم است. برای اطلاعات بیشتر می توانید به [۳، ۱۵] مراجعه کنید.

۳

## ۳  الگوریتم مجموعه عضوی انعکاس آفین (SM-AP)

برای معرفی الگوریتم SM-AP ابتدا ماتریس ورودی $\mathbf{X}(k)$، بردار خطا $\mathbf{e}(k)$، بردار مرجع $\mathbf{d}(k)$ بردار نویز $\mathbf{n}(k)$ و بردار محدودیت $\boldsymbol{\gamma}(k)$ را مشابه زیر تعریف می کنیم:

$$\begin{aligned}
\mathbf{X}(k) &= [\mathbf{x}(k)\ \mathbf{x}(k-1)\ \cdots\ \mathbf{x}(k-L)] \in \mathbb{R}^{(N+1)\times(L+1)}, \\
\mathbf{e}(k) &= [e(k)\ \epsilon(k-1)\ ...\ \epsilon(k-L)]^T \in \mathbb{R}^{L+1}, \\
\mathbf{d}(k) &= [d(k)\ d(k-1)\ \cdots\ d(k-L)]^T \in \mathbb{R}^{L+1}, \\
\mathbf{n}(k) &= [n(k)\ n(k-1)\ ...\ n(k-L)]^T \in \mathbb{R}^{L+1}, \\
\boldsymbol{\gamma}(k) &= [\gamma_0(k)\ \gamma_1(k)\ \cdots\ \gamma_L(k)]^T \in \mathbb{R}^{L+1},
\end{aligned} \quad (4)$$

که $N$ مرتبه فیلتر تطبیقی است و $L$ پارامتر مرور داده می باشد، یعنی $L$ داده نهایی پیش از تکرار $k$ مورد استفاده قرار می گیرند. بردار خطا با $\mathbf{e}(k) \triangleq \mathbf{d}(k) - \mathbf{X}^T(k)\mathbf{w}(k)$ تعریف می شود و مولفه های بردار محدودیت باید در شرط $|\gamma_i(k)| \leq \bar{\gamma}$ برای $i = 0, \ldots, L$ صدق کند، که $\bar{\gamma} \in \mathbb{R}_+$ کران بالا برای بزرگی سیگنال خطا $e(k)$ می باشد.
الگوریتم SM-AP با معادله بازگشتی زیر ارائه می شود [۲]،

$$\mathbf{w}(k+1) = \begin{cases} \mathbf{w}(k) + \mathbf{X}(k)\mathbf{A}(k)(\mathbf{e}(k) - \boldsymbol{\gamma}(k)) & \text{اگر } |e(k)| > \bar{\gamma}, \\ \mathbf{w}(k) & \text{در غیر این صورت} \end{cases} \quad (5)$$

که در آن فرض می کنیم $\mathbf{A}(k) \triangleq (\mathbf{X}^T(k)\mathbf{X}(k))^{-1} \in \mathbb{R}^{L+1\times L+1}$ موجود می باشد، یعنی $\mathbf{X}^T(k)\mathbf{X}(k)$ وارون پذیر می باشد.

## ۴  مقاومت الگوریتم SM-AP

یک سناریوی شناسایی سیستم را در نظر بگیرید که در آن سیستم ناشناخته توسط $\mathbf{w}_o \in \mathbb{R}^{N+1}$ و سیگنال مرجع به گونه زیر تعریف شده است

$$\mathbf{d}(k) \triangleq \mathbf{X}^T(k)\mathbf{w}_o + \mathbf{n}(k). \quad (6)$$

که در آن $n(k) \in \mathbb{R}$ سیگنال نویز را نشان می دهد. چون هدف ما مطالعه خاصیت مقاومت الگوریتم است، تعریف زیر از $\widetilde{\mathbf{w}}(k) \in \mathbb{R}^{N+1}$ مفید خواهد بود

$$\widetilde{\mathbf{w}}(k) = \mathbf{w}_o - \mathbf{w}(k), \quad (7)$$

۴

که $\widetilde{\mathbf{w}}(k)$ نشان‌دهنده تفاوت بین $\mathbf{w}_o$ و بردار $\mathbf{w}(k)$ است که به دنبال تخمین آن هستیم. بنابراین سیگنال خطا به گونه زیر بازنویسی خواهد شد

$$\mathbf{e}(k) = \mathbf{X}^T(k)\mathbf{w}_o + \mathbf{n}(k) - \mathbf{X}^T(k)\mathbf{w}(k) = \underbrace{\mathbf{X}^T(k)\widetilde{\mathbf{w}}(k)}_{\triangleq \widetilde{\mathbf{e}}(k)} + \mathbf{n}(k), \quad (8)$$

که $\widetilde{\mathbf{e}}(k)$ خطای بدون نویز می‌باشد.

یکی از مشکلات اساسی در تحلیل مقاومت الگوریتم SM-AP وضعیت شرطی معادله (۵) می‌باشد. برای رفع این مشکل تابع مشخصه $f : R \times R_+ \rightarrow \{0, 1\}$ را به گونه زیر تعریف کنیم

$$f(e(k), \overline{\gamma}) = \begin{cases} 1 & \text{اگر } ||e(k)| > \overline{\gamma}, \\ 0 & \text{در غیر اینصورت}. \end{cases} \quad (9)$$

بدین ترتیب، معادله بازگشتی الگوریتم SM-AP به صورت زیر بازنویسی می‌شود

$$\mathbf{w}(k+1) = \mathbf{w}(k) + \mathbf{X}(k)\mathbf{A}(k)(\mathbf{e}(k) - \boldsymbol{\gamma}(k))f(e(k), \overline{\gamma}). \quad (10)$$

بعد از تفاضل $\mathbf{w}_o$ از دو طرف معادله (۱۰) خواهیم داشت

$$\widetilde{\mathbf{w}}(k+1) = \widetilde{\mathbf{w}}(k) - \mathbf{X}(k)\mathbf{A}(k)(\mathbf{e}(k) - \boldsymbol{\gamma}(k))f(e(k), \overline{\gamma}). \quad (11)$$

توجه داشته باشید که ماتریس $\mathbf{A}(k)$ معین مثبت متقارن است. برای آسانی در نمادگذاری اندیس k، و ارگومان‌های تابع f را از سمت راست آن حذف می‌کنیم. سپس $e(k)$ را مطابق (۸) تجزیه می‌کنیم و خواهیم داشت

$$\widetilde{\mathbf{w}}(k+1) = \widetilde{\mathbf{w}} - \mathbf{X}\mathbf{A}\widetilde{\mathbf{e}}f - \mathbf{X}\mathbf{A}\mathbf{n}f + \mathbf{X}\mathbf{A}\boldsymbol{\gamma}f, \quad (12)$$

از اینجا قضیه ۱ را بدست می‌آوریم.

قضیه ۱ (مقاومت موضعی الگوریتم SM-AP): برای الگوریتم SM-AP همواره داریم

$$||\widetilde{\mathbf{w}}(k+1)||^2 = ||\widetilde{\mathbf{w}}(k)||^2, \text{ اگر } f(e(k), \overline{\gamma}) = 0 \quad (13)$$

در غیر این صورت

$$\begin{cases} \frac{||\widetilde{\mathbf{w}}(k+1)||^2 + \widetilde{\mathbf{e}}^T\mathbf{A}\widetilde{\mathbf{e}}}{||\widetilde{\mathbf{w}}(k)||^2 + \mathbf{n}^T\mathbf{A}\mathbf{n}} < 1, & \text{اگر } \boldsymbol{\gamma}^T\mathbf{A}\boldsymbol{\gamma} < 2\boldsymbol{\gamma}^T\mathbf{A}\mathbf{n} \\ \frac{||\widetilde{\mathbf{w}}(k+1)||^2 + \widetilde{\mathbf{e}}^T\mathbf{A}\widetilde{\mathbf{e}}}{||\widetilde{\mathbf{w}}(k)||^2 + \mathbf{n}^T\mathbf{A}\mathbf{n}} = 1, & \text{اگر } \boldsymbol{\gamma}^T\mathbf{A}\boldsymbol{\gamma} = 2\boldsymbol{\gamma}^T\mathbf{A}\mathbf{n} \\ \frac{||\widetilde{\mathbf{w}}(k+1)||^2 + \widetilde{\mathbf{e}}^T\mathbf{A}\widetilde{\mathbf{e}}}{||\widetilde{\mathbf{w}}(k)||^2 + \mathbf{n}^T\mathbf{A}\mathbf{n}} > 1, & \text{اگر } \boldsymbol{\gamma}^T\mathbf{A}\boldsymbol{\gamma} > 2\boldsymbol{\gamma}^T\mathbf{A}\mathbf{n} \end{cases} \quad (14)$$

که اندیس k برای آسانی کار حذف شده است و فرض کرده ایم که $||\widetilde{w}(k)||^2 + n^T A n \neq 0$.

اثبات: ابتدا با بدست آوردن نرم افلیدسی دو طرف (۱۲) خواهیم داشت

$$\begin{aligned}||\widetilde{w}(k+1)||^2 =& \widetilde{w}^T\widetilde{w} - \widetilde{w}^T X A \widetilde{e} f - \widetilde{w}^T X A n f + \widetilde{w}^T X A \boldsymbol{\gamma} f - \widetilde{e}^T A^T X^T \widetilde{w} f \\ &+ \widetilde{e}^T A^T A^{-1} A \widetilde{e} f^2 + \widetilde{e}^T A^T A^{-1} A n f^2 - \widetilde{e}^T A^T A^{-1} A \boldsymbol{\gamma} f^2 \\ &- n^T A^T X^T \widetilde{w} f + n^T A^T A^{-1} A \widetilde{e} f^2 + n^T A^T A^{-1} A n f^2 \\ &- n^T A^T A^{-1} A \boldsymbol{\gamma} f^2 + \boldsymbol{\gamma}^T A^T X^T \widetilde{w} f - \boldsymbol{\gamma}^T A^T A^{-1} A \widetilde{e} f^2 \\ &- \boldsymbol{\gamma}^T A^T A^{-1} A n f^2 + \boldsymbol{\gamma}^T A^T A^{-1} A \boldsymbol{\gamma} f^2 \\ =& ||\widetilde{w}||^2 - \widetilde{e}^T A \widetilde{e} f - \widetilde{e}^T A n f + \widetilde{e}^T A \boldsymbol{\gamma} f - \widetilde{e}^T A \widetilde{e} f + \widetilde{e}^T A \widetilde{e} f^2 + \widetilde{e}^T A n f^2 \\ &- \widetilde{e}^T A \boldsymbol{\gamma} f^2 - n^T A \widetilde{e} f + n^T A \widetilde{e} f^2 + n^T A n f^2 - n^T A \boldsymbol{\gamma} f^2 + \boldsymbol{\gamma}^T A \widetilde{e} f \\ &- \boldsymbol{\gamma}^T A \widetilde{e} f^2 - \boldsymbol{\gamma}^T A n f^2 + \boldsymbol{\gamma}^T A \boldsymbol{\gamma} f^2 \,, \quad (15)\end{aligned}$$

که در آن $\widetilde{e}(k) = X^T(k)\widetilde{w}(k)$ و $A^{-1} = X^T(k)X(k)$. از معادله بالا مشاهده می کنیم که برای $f = 0$ داریم

$$||\widetilde{w}(k+1)||^2 = ||\widetilde{w}(k)||^2 \quad (16)$$

و این نتیجه قابل انتظار است زیرا $f = 0$ یعنی هیچ بروزرسانی صورت نگرفته است. اما زمانی که $f = 1$ است رابطه زیر را از (۱۵) خواهیم داشت

$$||\widetilde{w}(k+1)||^2 = ||\widetilde{w}||^2 - \widetilde{e}^T A \widetilde{e} + n^T A n - 2\boldsymbol{\gamma}^T A n + \boldsymbol{\gamma}^T A \boldsymbol{\gamma} \,. \quad (17)$$

بعد مرتب سازی عبارت های معادله بالا خواهیم داشت

$$||\widetilde{w}(k+1)||^2 + \widetilde{e}^T A \widetilde{e} = ||\widetilde{w}||^2 + n^T A n - 2\boldsymbol{\gamma}^T A n + \boldsymbol{\gamma}^T A \boldsymbol{\gamma} \,. \quad (18)$$

بنابراین داریم

$$\begin{aligned}||\widetilde{w}(k+1)||^2 + \widetilde{e}^T A \widetilde{e} < ||\widetilde{w}||^2 + n^T A n \text{ اگر } \boldsymbol{\gamma}^T A \boldsymbol{\gamma} < 2\boldsymbol{\gamma}^T A n, \\ ||\widetilde{w}(k+1)||^2 + \widetilde{e}^T A \widetilde{e} = ||\widetilde{w}||^2 + n^T A n \text{ اگر } \boldsymbol{\gamma}^T A \boldsymbol{\gamma} = 2\boldsymbol{\gamma}^T A n, \quad (19) \\ ||\widetilde{w}(k+1)||^2 + \widetilde{e}^T A \widetilde{e} > ||\widetilde{w}||^2 + n^T A n \text{ اگر } \boldsymbol{\gamma}^T A \boldsymbol{\gamma} > 2\boldsymbol{\gamma}^T A n.\end{aligned}$$

فرض کنید $||\widetilde{w}||^2 + n^T A n \neq 0$، آن گاه مغادلات بالا را می توان به شکل زیر خلاصه کرد

$$\begin{cases} \dfrac{||\widetilde{w}(k+1)||^2 + \widetilde{e}^T A \widetilde{e}}{||\widetilde{w}(k)||^2 + n^T A n} < 1, & \text{اگر } \boldsymbol{\gamma}^T A \boldsymbol{\gamma} < 2\boldsymbol{\gamma}^T A n \\ \dfrac{||\widetilde{w}(k+1)||^2 + \widetilde{e}^T A \widetilde{e}}{||\widetilde{w}(k)||^2 + n^T A n} = 1, & \text{اگر } \boldsymbol{\gamma}^T A \boldsymbol{\gamma} = 2\boldsymbol{\gamma}^T A n \quad . \quad \square \quad (20) \\ \dfrac{||\widetilde{w}(k+1)||^2 + \widetilde{e}^T A \widetilde{e}}{||\widetilde{w}(k)||^2 + n^T A n} > 1, & \text{اگر } \boldsymbol{\gamma}^T A \boldsymbol{\gamma} > 2\boldsymbol{\gamma}^T A n \end{cases}$$

ترکیب دو نامساوی اول (۱۴) که مطابق حالت های $\boldsymbol{\gamma}^T\mathbf{A}\boldsymbol{\gamma} \leq 2\boldsymbol{\gamma}^T\mathbf{A}\mathbf{n}$ است، نتیجه جالبی را نشان خواهد داد. در واقع بیان خواهد کرد که برای هر بردار محدودیت $\boldsymbol{\gamma}$ که در این شرط صدق کند فارغ از نوع $\mathbf{n}(k)$ خواهیم داشت

$$||\widetilde{\mathbf{w}}(k+1)||^2 + \widetilde{\mathbf{e}}^T\mathbf{A}\widetilde{\mathbf{e}} \leq ||\widetilde{\mathbf{w}}(k)||^2 + \mathbf{n}^T\mathbf{A}\mathbf{n}, \qquad (21)$$

در نتیجه می توانیم خاصیت مقاومت سراسری را برای الگوریتم SM-AP ارائه دهیم.

تبصره ۱ (مقاومت سراسری الگوریتم SM-AP): فرض کنید که الگوریتم SM-AP از لحظه ۰ (مقداردهی اولیه) تا لحظه K با اختیار کردن $\boldsymbol{\gamma}$ به گونه ای که $\boldsymbol{\gamma}^T\mathbf{A}\boldsymbol{\gamma} \leq 2\boldsymbol{\gamma}^T\mathbf{A}\mathbf{n}$ صدق می کند اجرا می شود. آنگاه همواره خواهیم داشت

$$\frac{||\widetilde{\mathbf{w}}(K)||^2 + \sum_{k\in\mathcal{K}_{\text{up}}} \widetilde{\mathbf{e}}^T\mathbf{A}\widetilde{\mathbf{e}}}{||\widetilde{\mathbf{w}}(0)||^2 + \sum_{k\in\mathcal{K}_{\text{up}}} \mathbf{n}^T\mathbf{A}\mathbf{n}} \leq 1, \qquad (22)$$

که $\mathcal{K}_{\text{up}} \neq \emptyset$ مجموعه تکرارهایی است که بروزرسانی $\mathbf{w}(k)$ صورت گرفته است و تساوی زمانی برقرار است که $\boldsymbol{\gamma}^T\mathbf{A}\boldsymbol{\gamma} = 2\boldsymbol{\gamma}^T\mathbf{A}\mathbf{n}$ برای هر $k \in \mathcal{K}_{\text{up}}$. اگر $\mathcal{K}_{\text{up}} = \emptyset$، آن گاه $||\widetilde{\mathbf{w}}(K)||^2 = ||\widetilde{\mathbf{w}}(0)||^2$، یعنی حالتی که هیچ گونه بروزرسانی صورت نگرفته است.

اثبات: مجموعه همه تکرارهای تحت آنالیز را با $\mathcal{K} \triangleq \{0, 1, 2, \ldots, K-1\}$ نمایش دهید. فرض کنید $\mathcal{K}_{\text{up}} \subseteq \mathcal{K}$ شامل تکرارهایی است که در آنها بروزرسانی صورت گرفته است و $\mathcal{K}_{\text{up}}^{\text{c}} \triangleq \mathcal{K} \setminus \mathcal{K}_{\text{up}}$ مجموعه تکرارهایی است که در آنها بروزرسانی صورت نگرفته است. از قضیه ۱ هنگامی که بروزرسانی انجام شده است، نامساوی (۲۱) برقرار است به شرطی که $\boldsymbol{\gamma}$ به گونه ای انتخاب شده باشد که $\boldsymbol{\gamma}^T\mathbf{A}\boldsymbol{\gamma} \leq 2\boldsymbol{\gamma}^T\mathbf{A}\mathbf{n}$ صدق می کند. در این صورت، با جمع کردن این نامساوی برای همه $k \in \mathcal{K}_{\text{up}}$ خواهیم داشت

$$\sum_{k\in\mathcal{K}_{\text{up}}} \left(||\widetilde{\mathbf{w}}(k+1)||^2 + \widetilde{\mathbf{e}}^T\mathbf{A}\widetilde{\mathbf{e}}\right) \leq \sum_{k\in\mathcal{K}_{\text{up}}} \left(||\widetilde{\mathbf{w}}(k)||^2 + \mathbf{n}^T\mathbf{A}\mathbf{n}\right). \qquad (23)$$

توجه داشته باشید که $\boldsymbol{\gamma}$، $\widetilde{\mathbf{e}}$، $\mathbf{n}$ و $\mathbf{A}$ وابسته به $k$ هستند که برای آسانی کار حذف شده است. به علاوه برای تکرارهایی که بروزرسانی صورت نگرفته است رابطه (۱۳) را داریم، که می تواند برای همه $k \in \mathcal{K}_{\text{up}}^{\text{c}}$ جمع بسته شود، پس داریم

$$\sum_{k\in\mathcal{K}_{\text{up}}^{\text{c}}} ||\widetilde{\mathbf{w}}(k+1)||^2 = \sum_{k\in\mathcal{K}_{\text{up}}^{\text{c}}} ||\widetilde{\mathbf{w}}(k)||^2. \qquad (24)$$

۷

با جمع بستن (۲۳) و (۲۴) خواهیم داشت

$$\sum_{k \in \mathcal{K}} ||\widetilde{\mathbf{w}}(k+1)||^2 + \sum_{k \in \mathcal{K}_{up}} \widetilde{\mathbf{e}}^T \mathbf{A} \widetilde{\mathbf{e}} \leq \sum_{k \in \mathcal{K}} ||\widetilde{\mathbf{w}}(k)||^2 + \sum_{k \in \mathcal{K}_{up}} \mathbf{n}^T \mathbf{A} \mathbf{n}. \quad (۲۵)$$

اما مقادیر زیادی از $||\widetilde{\mathbf{w}}(k)||^2$ از دو طرف معادله بالا حذف می شوند و خواهیم داشت

$$||\widetilde{\mathbf{w}}(K)||^2 + \sum_{k \in \mathcal{K}_{up}} \widetilde{\mathbf{e}}^T \mathbf{A} \widetilde{\mathbf{e}} \leq ||\widetilde{\mathbf{w}}(0)||^2 + \sum_{k \in \mathcal{K}_{up}} \mathbf{n}^T \mathbf{A} \mathbf{n}. \quad (۲۶)$$

با فرض غیر صفر بودت مخرج خواهیم داشت

$$\frac{||\widetilde{\mathbf{w}}(K)||^2 + \sum_{k \in \mathcal{K}_{up}} \widetilde{\mathbf{e}}^T \mathbf{A} \widetilde{\mathbf{e}}}{||\widetilde{\mathbf{w}}(0)||^2 + \sum_{k \in \mathcal{K}_{up}} \mathbf{n}^T \mathbf{A} \mathbf{n}} \leq 1. \quad (۲۷)$$

این رابطه برای همه K ها برقرار است به شرطی که $\boldsymbol{\gamma}^T \mathbf{A} \boldsymbol{\gamma} \leq 2\boldsymbol{\gamma}^T \mathbf{A} \mathbf{n}$ برای تکرارهایی که بروزرسانی صورت می گیرد صدق کند. $\square$

توجه داشته باشید که در الگوریتم SM-AP شرط $\boldsymbol{\gamma}^T \mathbf{A} \boldsymbol{\gamma} \leq 2\boldsymbol{\gamma}^T \mathbf{A} \mathbf{n}$ باید برقرار باشد تا مقاومت با نرم $l_2$ از پارامترهای نااطمینانی $\{\widetilde{\mathbf{w}}(0), \{\mathbf{n}(k)\}_{0 \leq k \leq K}\}$ به پارامترهای خطا $\{\widetilde{\mathbf{w}}(K), \{\widetilde{e}(k)\}_{0 \leq k \leq K}\}$ حفظ شود. اما سوال بعدی این است که آیا بردار محدودیت $\boldsymbol{\gamma}$ وجود دارد که ایت شرط همواره برقرار باشد؟ تبصره ۲ به این سوال جواب می دهد و مثالی برای $\boldsymbol{\gamma}$ ارائه می دهد.

تبصره ۲: فرض کنید که بردار محدودیت $\boldsymbol{\gamma}(k) = c\mathbf{n}(k)$ برای الگوریتم SM-AP داده شده است که $\mathbf{n}(k)$ بردار نویز است و مطابق (۶) تعریف شده است. اگر $0 \leq c \leq 2$، آن گاه شرط $\boldsymbol{\gamma}^T \mathbf{A} \boldsymbol{\gamma} \leq 2\boldsymbol{\gamma}^T \mathbf{A} \mathbf{n}$ همواره برقرار است. در نتیجه الگوریتم SM-AP بر اساس تبصره ۱ به طور سراسری مقاوم می شود.

اثبات: در $\boldsymbol{\gamma}^T \mathbf{A} \boldsymbol{\gamma} \leq 2\boldsymbol{\gamma}^T \mathbf{A} \mathbf{n}$، $\boldsymbol{\gamma}(k) = c\mathbf{n}(k)$ قرار دهید ، در نتیجه خواهیم داشت $(c^2 - 2c)\mathbf{n}^T(k)\mathbf{A}(k)\mathbf{n}(k) \leq 0$، که چون $\mathbf{A}(k)$ ماتریس مثبت است برای

$$c^2 - 2c \leq 0 \Rightarrow 0 \leq c \leq 2 \quad (۲۸)$$

برقرار است. در نتیجه با تبصره ۱، برای $\boldsymbol{\gamma}(k) = c\mathbf{n}(k)$ ارائه شده الگوریتم SM-AP به طور سراسری مقاوم است. $\square$

توجه داشته باشید که $\boldsymbol{\gamma}(k)$ ارائه شده در تبصره ۲ عملاً قابل دستیابی نیست زیرا $\mathbf{n}(k)$ قابل دستیابی نیست. در نتیجه تبصره ۲ بیانگر وجود $\boldsymbol{\gamma}(k)$ ای است که در شرط $\boldsymbol{\gamma}^T \mathbf{A} \boldsymbol{\gamma} < 2\boldsymbol{\gamma}^T \mathbf{A} \mathbf{n}$

صدق می کند. اما در واقع تا کنون $\gamma(k)$ که عملاً قابل دستیابی باشد و در شرط $\gamma^T A \gamma \leq 2\gamma^T A n$ برای همه تکرارها صدق کند ارائه نشده است. حتی SC-CV [2]که یک انتخاب رایج برای بردار محدودیت است گاهی اوقات در این شرط صدق نمی کند. اما این بدین معنی نیست که الگوریتم SM-AP می تواند واگرا باشد. در واقع برای هر انتخاب $\gamma(k)$، الگوریتم SM-AP به هیچ وجه واگرا نمی شود.

وقتی که $\bar{\gamma} > |e(k)|$ و الگوریتم SM-AP بروزرسانی انجام می دهد، $w(k+1)$ را به عنوان جواب مسئله بهینه سازی زیر ارائه می دهد [2, 12]:

$$\text{مینیمم} \quad ||w(k+1) - w(k)||^2$$
$$\text{به شرط} \quad d(k) - X^T(k)w(k+1) = \gamma(k). \tag{29}$$

شرط این مسئله الزام می کند که خطای پسین $\epsilon(k-l) \triangleq d(k-l) - x^T(k-l)w(k+1)$ برابر $\gamma_l(k)$ متناظرش باشد، که همان طور که در ۳ بیان شده است توسط $\bar{\gamma}$ کراندار است. پس در نتیجه خواهیم داشت

$$|\epsilon(k-l)| = |d(k-l) - x^T(k-l)w(k+1)| \leq \bar{\gamma},$$
$$|x^T(k-l)\widetilde{w}(k+1) + n(k-l)| \leq \bar{\gamma}, \tag{30}$$

که برای همه تکرارها و پارامترها معتبر است. در نتیجه داریم

$$-\bar{\gamma} - n(k-l) \leq x^T(k-l)\widetilde{w}(k+1) \leq \bar{\gamma} - n(k-l). \tag{31}$$

چون نویز کراندار است و $\bar{\gamma} < \infty$، داریم

$$-\infty < \sum_{i=0}^{N} x_i(k-l)\tilde{w}_i(k+1) < \infty, \tag{32}$$

که $x_i(k-l), \tilde{w}_i(k+1) \in \mathbb{R}$ به ترتیب نشان دهنده $i$امین مولفه بردارهای $x(k-l), \widetilde{w}(k+1) \in \mathbb{R}^{N+1}$ هستند. در نتیجه $|\tilde{w}_i(k+1)|$ کراندار است و ایجاب می کند که $||\widetilde{w}(k+1)||^2 < \infty$ که بدین معنی است که زمانی که بردار محدودین به طور مناسب انتخاب نشده است الگوریتم SM-AP واگرا نمی شود. در بخش بعدی با اختیار کردن یک بردار محدودیت تصادفی نشان خواهیم داد که نتیجه الگوریتم SM-AP ضعیف می شود اما هیچ گاه این الگوریتم واگرا نمی شود.

## ۵  شبیه سازی

در این بخش الگوریتم SM-AP برای شناسایی سیستم به کار برده می شود. سیستم ناشناخته ۱۰ مولفه دارد که از توزیع نرمال استاندارد گرفته شده است. سیگنال نویز $n(k)$ دارای توزیع نرمال

۹

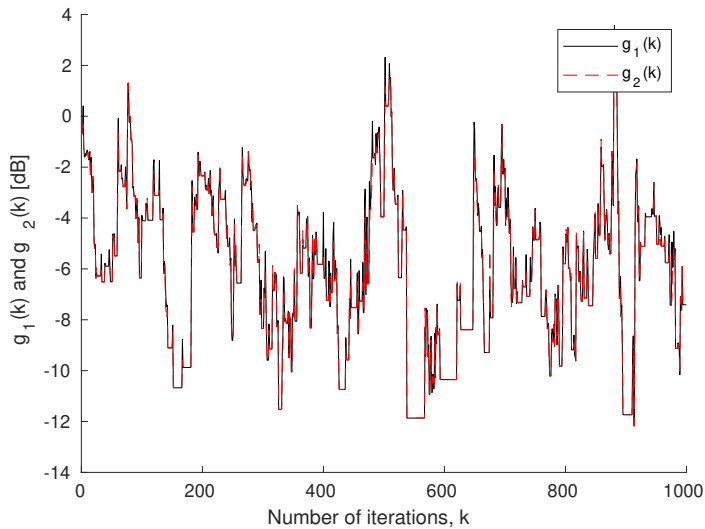

شکل ۱: زمانی که در الگوریتم SM-AP با در نظر گرفتن بردار محدودیت عمومی بروزرسانی صورت می گیرد مقادیر $g_1(k)$ و $g_2(k)$ مقدار صورت و مخرج در (۱۴) در قضیه ۱ می باشند، در غیر این صورت $g_1(k) = ||\widetilde{\mathbf{w}}(k+1)||^2$ و $g_2(k) = ||\widetilde{\mathbf{w}}(k)||^2$.

$\mathbf{w}(0) = [0 \cdots 0]^T \in \mathbb{R}^{10}$ استاندارد با میانگین صفر و واریانس $\sigma_n^2 = 0.01$ است. همچنین $\delta = 10^{-12}$ و $\overline{\gamma} = \sqrt{5\sigma_n^2} = 0.2236$ هستند. سیگنال ورودی با $x(k) = 0.95x(k-1) + n(k-1)$ تعریف شده است، و نسبت نویز به سیگنال برابر ۲۰ دسیبل است. پارامتر مرور داده $L = 2$ برگزیده شده است و سه نوع بردار محدودیت مورد استفاده قرار گرفته اند: بردار محدودیت عمومی، بردار محدودیت SC-CV و بردار محدودیت برابر بردار نویز. در بردار محدودیت عمومی $\gamma_l(k) = \overline{\gamma}$ برای $0 \leq l \leq L$، که حالتی را در نظر می گیرد که بردار محدودیت به طور مناسب برگزیده نشده است. بردار محدودیت SC-CV به این گونه تعریف شده است که $\gamma_0(k) = \overline{\gamma}\frac{e(k)}{|e(k)|}$ و $\gamma_l(k) = \epsilon(k-l)$ برای $1 \leq l \leq L$. در نهایت بردار محدودیت برابر بردار نویز بدین گونه است که $\boldsymbol{\gamma}(k) = \mathbf{n}(k)$. وقتی که بروزرسانی صورت می گیرد صورت و مخرج قضیه ۱ را به ترتیب با $g_1(k)$ و $g_2(k)$ نشان می دهیم، در غیر این صورت $g_1(k) = ||\widetilde{\mathbf{w}}(k+1)||^2$ و $g_2(k) = ||\widetilde{\mathbf{w}}(k)||^2$.

نتایج ارائه شده در شکل ۱ نشان می دهد که زمانی که بردار محدودیت عمومی به کار گرفته شده است، تعداد تکرارهای زیادی وجود دارند که در آن ها $g_2(k) > g_1(k)$ (حدود ۲۹۳ از ۱۰۰۰ تکرار). این نتیجه قابل انتظار است زیرا بردار محدودیت عمومی به طور مستقیم یا غیر مستقیم سیگنال نویز را لحاظ قرار نمی دهد. بنابراین شرط مقاومت $\boldsymbol{\gamma}^T(k)\mathbf{A}(k)\boldsymbol{\gamma}(k) \leq 2\boldsymbol{\gamma}^T(k)\mathbf{A}(k)\mathbf{n}(k)$ مورد بررسی قرار نمی گیرد.

هنگامی که بردار محدودیت SC-CV در الگوریتم SM-AP مورد استفاده قرار می گیرد، تعداد تکرارهای بسیار کمتری وجود دارند که در آنها $g_2(k) > g_1(k)$ (۲۱ از ۱۰۰۰ تکرار)،

۱۰

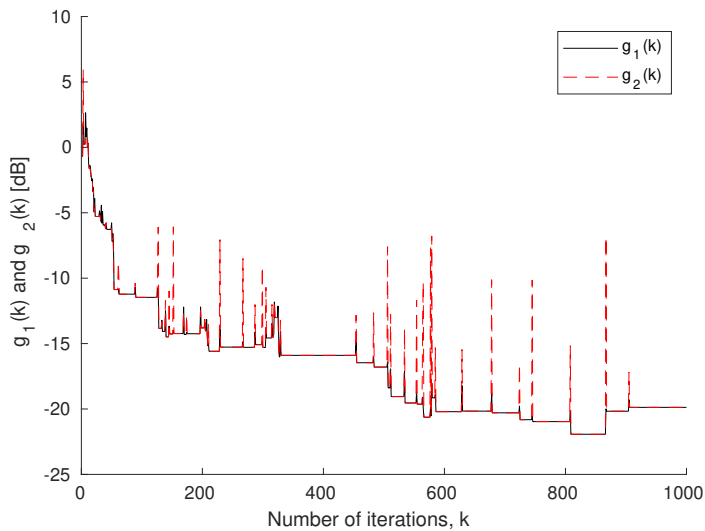

شکل ۲: زمانی که در الگوریتم SM-AP با در نظر گرفتن بردار محدودیت SC-CV بروزرسانی صورت می گیرد مقادیر $g_1(k)$ و $g_2(k)$ مقدار صورت و مخرج در (۱۴) در قضیه ۱ می باشند، در غیر این صورت $g_1(k) = ||\widetilde{\mathbf{w}}(k+1)||^2$ و $g_2(k) = ||\widetilde{\mathbf{w}}(k)||^2$.

همان طور که در شکل ۲ نشان داده شده است. این بیان می کند که ردار محدودیت SC-CV که بسیار رایج است نیز مقاومت سراسری الگوریتم SM-AP را تضمین نمی کند.

شکل ۳ نتایج را برای زمانی نشان می دهد که بردار محدودیت $\boldsymbol{\gamma}(k) = \mathbf{n}(k)$ برگزیده شده است. در این حالت مشاهده می کنیم که همواره برای هر k داریم $g_2(k) \leq g_1(k)$، و این تایید کننده تبصره ۳ است. به عبارت دیگر این بردار محدودیت مقاومت سراسری الگوریتم SM-AP را تضمین می کند.

شکل ۴ دنباله $\{||\widetilde{\mathbf{w}}(k)||^2\}$ یرای الگوریتم های AP و SM-AP نشان می دهد. برای الگوریتم AP مقدار گام بروزرسانی $\mu$ ۰/۹ و ۰/۰۵ انتخاب شده است. برای الگوریتم SM-AP سه ردار محدودیت از پیش تعیین شده مورد استفاده قرار گرفته است. در مورد الگوریتم AP مشاهده می کنیم که رفتار $\{||\widetilde{\mathbf{w}}(k)||^2\}$ به شدت نامنظم است. حتی زمانی که $\mu$ کوچک برگزیده شده است، تعداد تکرارهای زیادی مشاهده می کنیم که $||\widetilde{\mathbf{w}}(k+1)||^2 > ||\widetilde{\mathbf{w}}(k)||^2$ (۴۳۱ از ۱۰۰۰ تکرار). الگوریتم SM-AP زمانی که بردار محدودیت عمومی رفتاری بسیار مشابه الگوریتم AP دارد هنگامی که $\mu$ انتخاب شده است. اما زمانی که بردار محدودیت SC-CV استفاده شده است، مشاهده می کنیم که تعداد حالت های $||\widetilde{\mathbf{w}}(k+1)||^2 > ||\widetilde{\mathbf{w}}(k)||^2$ به شدت کاهش یافته است (۲۹ از ۱۰۰۰ تکرار). همچنین زمانی که بردار محدودیت $\boldsymbol{\gamma}(k) = \mathbf{n}(k)$ مورد استفاده قرار گرفته است، مشاهده می کنیم که دنباله $\{||\widetilde{\mathbf{w}}(k)||^2\}$ نزولی است.

میانگین مربعات خطا برای لگوریتم های AP و SM-AP در شکل ۵ نمایش داده شده

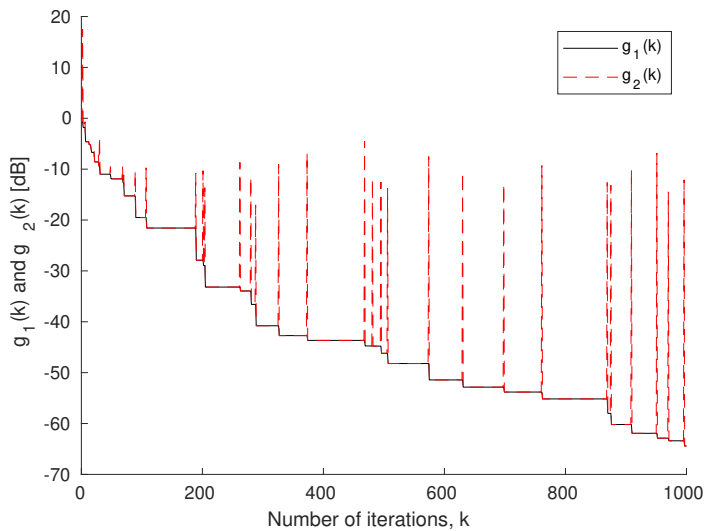

شکل ۳: زمانی که در الگوریتم SM-AP با در نظر گرفتن بردار محدودیت $\boldsymbol{\gamma}(k) = \mathbf{n}(k)$ بروزرسانی صورت می گیرد مقادیر $g_1(k)$ و $g_2(k)$ مقدار صورت و مخرج در (۱۴) در قضیه ۱ می باشند، در غیر این صورت $g_1(k) = ||\widetilde{\mathbf{w}}(k+1)||^2$ و $g_2(k) = ||\widetilde{\mathbf{w}}(k)||^2$.

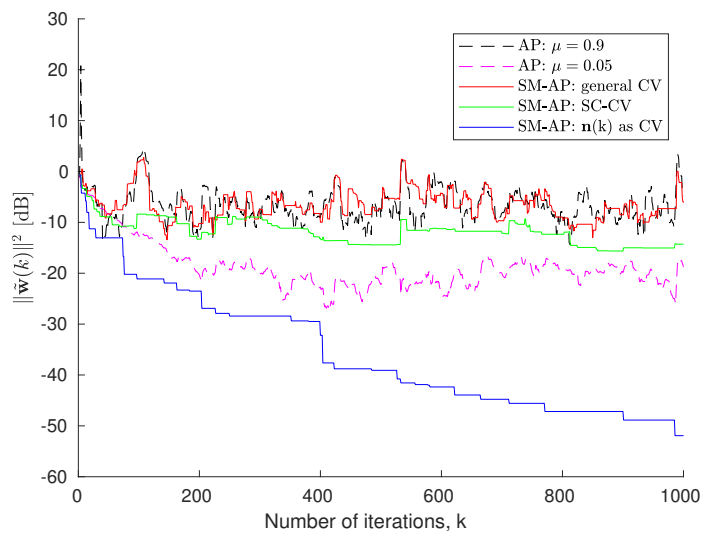

شکل ۴: $||\widetilde{\mathbf{w}}(k)||^2 \triangleq ||\mathbf{w}(k) - \mathbf{w}_o||^2$ برای الگوریتم های SM-AP و AP.

است. این با گرفتن میانگین از ۱۰۰۰ اجرای مستقل به دست آمده است. مشاهده می کنید که در الگوریتم AP تفاوت بین سرعت همگرایی و سطح میانگین مربعات خطا فاحش است. با نا دیده گرفتن الگوریتم SM-AP در حالتی که بردار محدودیت عمومی استفاده شده است (که

۱۲

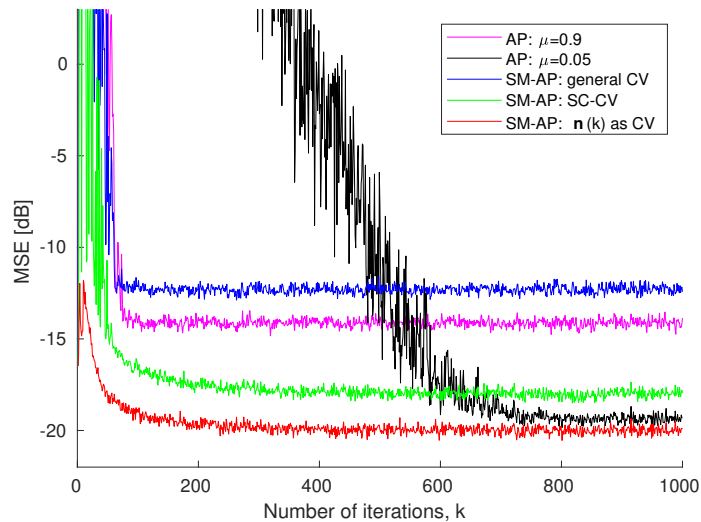

شکل ۵: منحنی های میانگین مربعات خطا برای الگوریتم های AP و SM-AP با در نظر گرفتن بردار محدودیت های متفاوت.

حالتی نا مرسوم است) الگوریتم AP نمی تواند به طور هم زمان سرعت همگرایی بالا و سطح میانگین مربعات خطا پایین داشته باشد، اما الگوریتم SM-AP قادر به کسب چنین نتیجه ای است. علاوه بر این مشاهده می کنید زمانی که $\gamma(k) = n(k)$ بهترین نتیجه را برای سرعت همگرایی و سطح میانگین مربعات خطا داریم، اما بردار محدودیت SC-CV نتیجه ای بسیار مشابه را نتیجه می دهد. میانگین تعداد دفعات بروزرسانی در الگوریتم SM-AP با در نظر گرفتن بردار محدودیت عمومی، SC-CV و نویز به ترتیب برابر ۳۴/۷%، ۹/۸% و ۳/۷% است. توجه داشته باشید حتی زمانی که بردار محدودیت عمومی به کار گرفته می شود، الگوریتم SM-AP همگرا می شود، اگرچه عملکرد ضعیفی را از خود نشان می دهد.

## ۶  نتایج

در این مقاله مقاومت الگوریتم SM-AP بررسی شد. در ابتدا مقاومت موضعی مورد بررسی قرار گرفت و سپس به کمک آن مقاومت سراسری بررسی شد. علاوه بر این نشان دادیم که الگوریتم SM-AP فارغ از نحوه انتخاب پارامترهای الگوریتم، به هیچ وجه واگرا نمی شود. همچنین در این مقاله نشان دادیم که مقاومت الگوریتم SM-AP به شدت به انتخاب بردار محدودیت وابسته است. در نهایت صحت اظهارات ارائه شده را با شبیه سازی بررسی کردیم.

۱۳

# مراجع